\documentclass[12pt]{article}
\newcommand {\citeAY} [1] {\citeNP {#1}}%
\newcommand {\citeAPY}[1] {\citeN  {#1}}%
\usepackage{xchicago}
\renewcommand {\showoriginalref}[1]{}
\renewcommand {\showCODEN}[1]{}
\renewcommand {\showISSN}[1]{}
\renewcommand {\showMR}[3]{}
\def\theequation{\thesection.\arabic{equation}}
\newcommand\eq[1] {(\ref{#1})}

\newcommand{\nonum}{\nonumber \\}
\newcommand{\beqa}{\begin{eqnarray}}
\newcommand{\eeqa}[1]{\label{#1}\end{eqnarray}}
\newcommand{\beq}{\begin{equation}}
\newcommand{\eeq}[1]{\label{#1}\end{equation}}
\newcommand{\Grad}{\nabla}
\newcommand{\Div}{\nabla \cdot}
\newcommand{\Curl}{\nabla \times}

\newcommand{\Gd}{\delta}

\newcommand{\Gve}{\varepsilon}

\newcommand{\Gk}{\kappa}

\newcommand{\Gm}{\mu}

\newcommand{\Gt}{\theta}

\newcommand{\Gs}{\sigma}

\newcommand{\Go}{\omega}

\newcommand{\GO}{\Omega}

\newcommand{\CC}{{\cal C}}
\newcommand{\CD}{{\cal D}}
\newcommand{\CE}{{\cal E}}

\newcommand{\CG}{{\cal G}}
\newcommand{\CH}{{\cal H}}

\newcommand{\CK}{{\cal K}}
\newcommand{\CL}{{\cal L}}
\newcommand{\CM}{{\cal M}}

\newcommand{\CP}{{\cal P}}
\newcommand{\CQ}{{\cal Q}}
\newcommand{\CR}{{\cal R}}

\usepackage{graphics}
\usepackage{amssymb}
\def\half{{\scriptstyle\frac{1}{2}}}
\def\ep{e^\prime}
\def\epp{e^{\prime\prime}}
\def\fp{f^\prime}
\def\fpp{f^{\prime\prime}}
\def\up{u^\prime}
\def\upp{u^{\prime\prime}}
\def\Cp{C^\prime}
\def\Cpp{C^{\prime\prime}}
\def\cp{c^\prime}
\def\cpp{c^{\prime\prime}}
\def\rp{\rho^\prime}
\def\rpp{\rho^{\prime\prime}}
\def\sp{\sigma^\prime}
\def\spp{\sigma^{\prime\prime}}
\def\pp{p^\prime}
\def\ppp{p^{\prime\prime}}

\def\tpp{\tau^{\prime\prime}}
\def\np{\nu^\prime}

\def\etpp{\eta^{\prime\prime}}
\def\pip{\pi^\prime}

\def\calC{{\cal C}}
\def\calP{{\cal P}}

\def\calQ{{\cal Q}}
\def\calD{{\cal D}}

\begin{document}
\vspace{-1in}
\title{ Minimum variational principles for time-harmonic waves
in a dissipative medium and associated variational principles
of Hashin--Shtrikman type}
\author{Graeme W. Milton\\
\small{Department of Mathematics, University of Utah, Salt Lake City UT 84112, USA}\\ \\
John R. Willis\\
\small{University of Cambridge,}\\
\small{Department of Applied Mathematics and
Theoretical Physics,}\\
\small{Wilberforce Road, Cambridge CB3 0WA, U.K.}}
\date{}
\maketitle
\begin{abstract}
\vskip2mm
Minimization variational principles for linear elastodynamic, acoustic, or electromagnetic time-harmonic waves in dissipative media
were obtained by Milton, Seppecher and Bouchitt\'e generalizing the quasistatic
variational principles of Cherkaev and Gibiansky. Here a further generalization is
made to allow for a much wider variety of boundary conditions, and in particular Dirichlet
and Neumann boundary conditions. In addition minimization or maximization principles of the
Hashin-Shtrikman type, incorporating ``polarization fields'', are developed.  

\end{abstract}
\section{Introduction}
Stationary variational principles for wave equations have been the subject
of much attention [see, for example,
\citeAPY{Oden:1983:VMT}, \citeAPY{Lazzari:2000:VPE} 
and \citeAPY{Altay:2004:FEC}, and references therein].
Among these the variational principles of \citeAPY{Gurtin:1964:VPL},
\citeAPY{Tao:1966:VPE} and Willis (\citeyearNP{Willis:1981:VPDP}, 
\citeyearNP{Willis:1984:VPOE}) are minimizing variational principles 
in the Laplace domain but not in the frequency domain. 
In the frequency domain in a dissipative medium they
correspond to saddle point variational principles 
(\citeAY{Borcea:1999:AAQ}, \citeAY{Milton:2009:MVP}).

Saddle point variational principles for time-harmonic waves in dissipative
media in the quasistatic limit, where the wavelengths and attenuation 
lengths are large compared to the size of the body, were derived by
\citeAPY{Cherkaev:1994:VPC}. They realized that by making a partial
Legendre transform one could convert them to minimization principles.
Equivalently, one could rewrite
the complex equations in a form involving the real and imaginary
parts of the fields with a real positive definite matrix entering
the consititutive law. From this reformulation, and from the differential
constraints on the fields, one immediately obtains the minimization principles.
Their variational principles were employed to give a simple
derivation (\citeAY{Milton:1990:CSP}) of existing
bounds on the effective complex electrical permittivity of lossy multiphase
composites. [These bounds, which generalized two-phase 
bounds of \citeAPY{Milton:1981:BCP} and \citeAPY{Bergman:1982:RBC},
were first
conjectured by \citeAPY{Golden:1985:BEP} and \citeAPY{Golden:1986:BCP}
and subsequently proved by \citeAPY{Bergman:1986:EDC}, \citeAPY{Milton:1987:MCEa},
and \citeAPY{Milton:1990:RCF}.]
Their variational principle also provided entirely new bounds
on the complex bulk and shear moduli of 
viscoelastic two-phase composites (\citeAY{Gibiansky:1993:EVM}; 
\citeAY{Gibiansky:1993:BCB}; 
\citeAY{Milton:1997:EVM}; \citeAY{Gibiansky:1997:BCB};
\citeAY{Gibiansky:1999:EVM}), which in turn led to bounds on the complex
thermal expansion coefficient of viscoelastic two-phase composites
(\citeAY{Berryman:2009:FDT}).

\citeAPY{Milton:2009:MVP} realized that the technique of 
Cherkaev and Gibiansky could be extended to obtain
minimization variational principles for time harmonic
waves in a body composed of dissipative material. In particular
they obtained minimization variational principles
for acoustics, elastodynamics and electromagnetism.
[In passing we mention that the final paragraph in that paper should be disregarded
as the inequality (6.8) does not provide any constraints on the moduli
inside the body as the difference between the right and left hand sides can
be expressed as the integral of a square.] An unappealing feature of these
variational principles is that they require boundary conditions on the 
fields which are difficult to physically impose, such as fixing the
real part of the displacement together with the real part of the traction
(for elastodynamics) or the real part of the tangential component
of $E$ together with the imaginary part of $H$ (for electromagnetism).

Here, by including what amounts to additional surface terms,
we derive more general variational principles which allow for all
sorts of boundary conditions.  These include the standard boundary
conditions where both real and imaginary parts of the displacement, 
traction, tangential component of $E$, or tangential component of
$H$ are specified. Like the variational principles of Milton,
Seppecher and Bouchitt\'e, these variational principles 
should be useful for tomography where one seeks information about the moduli inside a
body from a series of measurements of the fields at the surface of the body
from various different boundary conditions.   

We remark that the formulation in terms of minimization principles enables 
one to use the conjugate gradient algorithm for the numerical
solution using finite elements. This has been explored by \citeAPY{Richins:2010:NMS} 
for the Helmholtz equation, who also
obtain error bounds on the difference between the numerical solution and the exact solution as a function of the grid size.

Another completely different type of variational principle was derived by
Hashin and Shtrikman (\citeyearNP{Hashin:1962:VAT}, \citeyearNP{Hashin:1962:SVP}, 
\citeyearNP{Hashin:1962:VATE}, \citeyearNP{Hashin:1963:VAT}) to bound
the effective moduli of composites. This 
type of principle was extended to wave equations
as stationary variational principles by \citeAPY{Ben-Amoz:1966:VPA} and
Willis (\citeyearNP{Willis:1981:VPDP}, \citeyearNP{Willis:1984:VPOE}).
Here we derive Hashin--Shtrikman type minimization principles
for time harmonic waves in dissipative elastodynamic media. The
extension of these principles to acoustics and electrodynamics
exactly parallels the treatment given for elastodynamics
and is therefore omitted.

\setcounter{equation}{0}
\section{Preliminaries for elastodynamics}
\setcounter{equation}{0}
The propagation of stress waves in any body is governed by the equation of motion
\begin{equation}
{\rm div}\sigma + f = \dot p
\eeq{1}
where $\sigma$ denotes the stress tensor, $p$ is the momentum density (a vector) and $f$
is body force per unit volume. The body is characterized by its constitutive relations
which will be taken to have the linear form
\begin{equation}
\sigma = C*e,\;\;\;p = \rho*v
\eeq{2}
where the strain tensor $e$ and velocity vector $v$ are related to displacement $u$ by
\begin{equation}
e = \half\{\nabla u + (\nabla u)^T\},\;\;\;v = \dot u
\eeq{3}
The symbol $*$ denotes convolution with respect to time and thus the body is taken to be
viscoelastic. Although it is usually the case that the mass density $\rho$ is a real scalar,
the formulation is not complicated if $\rho$ is taken to be a tensor-valued convolution operator
and therefore this is assumed.

The case of time-harmonic disturbances will be assumed, in which case all fields become
functions of the spatial variable $x$ only, multiplied by the factor $e^{i\omega t}$
which will be suppressed in the sequel. In this case, the superposed dot denoting time derivative
is replaced by $i\omega$ and the convolution $*$ is replaced by ordinary multiplication.
If the body occupies a domain $\Omega$ and tractions $T_0=\sigma\cdot n$
are applied at its boundary $\partial\Omega$, then the rate of working of the surface tractions
and body force is
\begin{eqnarray}
{\dot W} &=& \int_{\partial\Omega}\,{\rm Re}(T_0e^{i\omega t})\cdot {\rm Re}(i\omega ue^{i\omega t})\,dS
 +\int_\Omega\,{\rm Re}(fe^{i\omega t})\cdot{\rm Re}(i\omega ue^{i\omega t})\,dx\nonumber\\
&=& \int_\Omega\,\{ {\rm Re}(\sigma e^{i\omega t}):{\rm Re}(i\omega e\,e^{i\omega t})
+{\rm Re}(i\omega p e^{i\omega t})\cdot{\rm Re}(i\omega u e^{i\omega t})\}\,dx,
\eeqa{4}
having employed the divergence theorem and the equation of motion \eq{1}. 
It follows, employing the time-reduced version of the constitutive relations \eq{2}, that the mean
rate of working of the applied forces is
\begin{equation}
\overline{\dot W} = \half \omega\int_\Omega\,\{\ep\Cpp\ep +\epp\Cpp\epp-\omega^2
(\up\rpp\up+\upp\rpp\upp)\}\,dx,
\eeq{5}
where $C=\Cp+i\Cpp$,
$e = \ep+i\epp$ etc. Also, the relevant contractions ($:$ or $\cdot$) are left implicit.
It follows, assuming that $\omega >0$, that positive-definiteness of
the quadratic forms $\ep\Cpp\ep$ and $\up(-\rpp)\up$ guarantees positive energy dissipation.
This will be assumed henceforth, though relaxation of these conditions (such as $\rpp=0$) will
be discussed after the main development is complete.

Now consider the time-harmonic problem, as described above, but with more general boundary conditions.
Three compatible boundary conditions are given at each point of $\partial\Omega$. These could be
for instance:
all three components of displacement; all three components of traction; two components of traction and the
orthogonal component of displacement (e.g. normal component of displacement and shear components
of traction). 

The problem with any such conditions can be described by the stationary principle
of \citeAPY{Willis:1984:VPOE}, whose time-reduced form is
\begin{equation}
\delta\int_\Omega\,\{\half eCe -\sigma_0 e -\half \omega^2 u\rho u - i\omega p_0 u\}\,dx =0,
\eeq{6}
the variation being taken over displacement fields $u$ that satisfy whatever conditions are prescribed
for displacements on $\partial\Omega$. The fields $\sigma_0$ and $p_0$ are not unique but are
any fields that satisfy the equation of motion \eq{1}, together with any conditions that are prescribed for the tractions on $\partial\Omega$. To confirm this claim, note that the variation is
\begin{equation}
\int_\Omega\,\{\delta e(\sigma-\sigma_0) - i\omega\delta u(p-p_0)\}\,dx
=\int_{\partial\Omega}\,\delta u(\sigma-\sigma_0)n\,dS -\int_\Omega\,\delta u[{\rm div}(\sigma-\sigma_0)
-i\omega(p-p_0)]\,dx.
\eeq{7}
Requiring the volume integral to vanish for all $\delta u$ implies the time-reduced
version of the equation of motion \eq{1}. The surface integral picks out only those
components of surface traction for which the corresponding component of displacement is not satisfied,
and hence the boundary conditions on traction are enforced.

There is a dual stationary principle, that
\begin{equation}
\delta\int_\Omega\,\{\half\sigma C^{-1}\sigma - e_0\sigma +\half p\rho^{-1} p - i\omega p u_0\}\,dx =0,
\eeq{8}
where $u_0$ is any displacement field that satisfies any given boundary conditions on displacement. The variation is taken over fields $\sigma$, $p$ that satisfy the equation of motion and any given traction
boundary conditions.

\section{Minimum variational principles for elastodynamics}
\setcounter{equation}{0}
First, the stationary principle \eq{6} is spelled out explicitly:
\begin{eqnarray}
&&\delta\int_\Omega\,\{\half(\ep\Cp\ep - \epp\Cp\epp -2\epp\Cpp\ep) -\ep\sp_0 +\epp\spp_0\nonumber\\
&&\hbox{\hskip 1in}
-\half\omega^2 (\up\rp\up - \upp\rp\upp -2\upp\rp\rp)+ \omega(\ppp_0\up+\pp_0\upp)\}\nonumber\\
&&+i[\half(\ep\Cpp\ep - \epp\Cpp\epp +2\epp\Cp\ep) -\epp\sp_0 -\ep\spp_0\nonumber\\
&&\hbox{\hskip 1in}
-\half\omega^2 (\up\rpp\up - \upp\rpp\upp +2\upp\rp\up)- \omega(\pp_0\up-\ppp_0\upp)]\}\,dx =0.
\eeqa{9}
Note that \eq{6} remains true if it is multiplied throughout by $e^{i\theta}$. Thus, the whole principle is delivered by requiring that
just the imaginary part is satisfied, but with $C$, $\rho$, $\sigma_0$, $p_0$ replaced by $Ce^{i\theta}$,
$\rho e^{i\theta}$  $\sigma_0e^{i\theta}$, $p_0e^{i\theta}$ for any $\theta$. Henceforth, just the imaginary part of \eq{9} will be
considered, while bearing in mind that $C$, $\rho$, $\sigma_0$ and $p_0$ can be changed as just indicated. For the validity of the reasoning to follow,
$\theta$ should be chosen such that the imaginary parts of $Ce^{i\theta}$, 
and $-\rho e^{i\theta}$ are
positive definite. If $\rho$ is real, as is typically the case, 
and $\Cpp$ is strictly positive definite,
such a replacement allows
the ensuing analysis to apply since (aside from in the variational
principles \eq{23} and \eq{25}) we assume that both $\Cpp$ and $-\rpp$ are 
strictly positive definite. 

The functional expressed in the imaginary part of \eq{9} is saddle-shaped. An alternative functional is obtained by performing
a partial duality transformation which involves $\sp$ and $\ppp$ in place of $\epp$ and $\omega\upp$. To clarify what is going on, note first that the time-reduced constitutive relations $\sigma = C e$, $p = i\omega \rho u$ imply that
\begin{equation}
\left(\matrix{\spp\cr
              \sp\cr}\right) = \left(\matrix{\Cpp & \Cp\cr
                                              \Cp & -\Cpp\cr}\right)
        \left(\matrix{\ep\cr
                      \epp\cr}\right),\;\;\;
\left(\matrix{\pp\cr
              -\ppp\cr}\right) = -\left(\matrix{\rpp & \rp\cr
                                                 \rp & -\rpp\cr}\right)
\left(\matrix{\omega\up\cr
              \omega\upp\cr}\right).
\label{9.5}\end{equation}
Now following \citeAPY{Cherkaev:1994:VPC},
elementary calculation gives
\begin{equation}
\half\left(\matrix{\ep &\epp\cr}\right)\left(\matrix{\Cpp & \Cp\cr
                       \Cp & -\Cpp\cr}\right)\left(\matrix{\ep\cr
                                                          \epp\cr}\right)
= \inf_{\sp}\left\{\sp\epp +\half\left(\matrix{\ep &\sp\cr}\right){\cal C}
\left(\matrix{\ep\cr
              \sp\cr}\right)\right\},
\eeq{10}
where
\begin{equation}
{\cal C} = \left(\matrix{\Cpp+\Cp(\Cpp)^{-1}\Cp & -\Cp(\Cpp)^{-1}\cr
                        -(\Cpp)^{-1}\Cp & (\Cpp)^{-1}\cr}\right)
\eeq{11}
is positive definite, as can be seen from the positivity of the associated quadratic
form
\beq \left(\matrix{\ep &\sp\cr}\right){\cal C}
\left(\matrix{\ep\cr 
              \sp\cr}\right)
=\ep\Cpp\ep+(\Cp\ep-\sp)(\Cpp)^{-1}(\Cp\ep-\sp).
\eeq{11.5}
The infimum is attained when $\sp$ satisfies (\ref{9.5})$_1$, or equivalently
$\sigma = C e$. Similarly,
\begin{equation}
\half\left(\matrix{\omega\up &\omega\upp\cr}\right)\left(\matrix{-\rpp & -\rp\cr
                       -\rp & \rpp\cr}\right)\left(\matrix{\omega\up\cr
                                                           \omega\upp\cr}\right)
= \inf_{\ppp}\left\{-\omega\upp\ppp +\half\left(\matrix{\omega\up &\ppp\cr}\right){\cal P}
\left(\matrix{\omega\up\cr
               \ppp\cr}\right)\right\},
\eeq{12}
where
\begin{equation}
{\cal P} = -\left(\matrix{\rpp+\rp(\rpp)^{-1}\rp & -\rp(\rpp)^{-1}\cr
                          -(\rpp)^{-1}\rp &   (\rpp)^{-1}\cr}\right)
\eeq{13}
is positive definite. Attainment of this infimum is consistent with (\ref{9.5})$_2$
or equivalently $p=i\omega\rho u$.

The imaginary part of \eq{9} can now be written
\begin{eqnarray}
&&\inf_{\up}\inf_{\sp,\ppp}\sup_{\upp}\int_\Omega\,\Biggl[\left(\matrix{\ep &\epp\cr}\right)
\left(\matrix{-\spp_0\cr
              \sp-\sp_0\cr}\right) -\left(\matrix{\omega\up &\omega\upp\cr}\right)
\left(\matrix{\pp_0\cr
              \ppp-\ppp_0\cr}\right)\nonumber\\
&&\hbox{\hskip 1in} +\half\left(\matrix{\ep &\sp\cr}\right){\cal C}
\left(\matrix{\ep\cr
              \sp\cr}\right)
+\half\left(\matrix{\omega\up & \ppp\cr}\right){\cal P}
\left(\matrix{\omega\up\cr
               \ppp\cr}\right)\Biggr].
\eeqa{14}

Now
\begin{equation}
\sup_{\upp}\int_\Omega\,\{\epp(\sp-\sp_0)-\omega\upp(\ppp-\ppp_0)\}\,dx
= \int_{\partial\Omega}\,\upp_0(\sp-\sp_0)n\,dS \equiv \int_\Omega\,
\{\epp_0(\sp-\sp_0)-\omega\upp_0(\ppp-\ppp_0)\}\,dx
\eeq{15}
so long as
\begin{equation}
{\rm div}\sp + f^\prime = -\omega\ppp \hbox{ in } \Omega
\eeq{16}
and $\sp$ satisfies any given traction conditions on $\partial\Omega$. Otherwise, the supremum is
infinite. Therefore, the variational principle \eq{14} reduces to
\begin{eqnarray}
&&\inf_{\up}\inf_{\sp}\int_\Omega\,\Biggl[\left(\matrix{\ep &\epp_0\cr}\right)
\left(\matrix{-\spp_0\cr
              \sp-\sp_0\cr}\right) - \left(\matrix{\omega\up &\omega\upp_0\cr}\right)
\left(\matrix{\pp_0\cr
              \ppp-\ppp_0\cr}\right)\nonumber\\
&&\hbox{\hskip 1in} +\half\left(\matrix{\ep &\sp\cr}\right){\cal C}
\left(\matrix{\ep\cr
              \sp\cr}\right)
+\half\left(\matrix{\omega\up & \ppp\cr}\right){\cal P}
\left(\matrix{\omega\up\cr
               \ppp\cr}\right)\Biggr]\,dx,
\eeqa{17}
where $\ppp$ is now given by \eq{16} and
the infimum is subject to any prescribed traction 
boundary conditions on $\sp$ and any prescribed displacement boundary conditions 
on $\up$. Since the products $\epp_0\sp_0$ and $\upp_0\ppp_0$ do not affect the minimizer
the variational principle further simplifies to
\beqa
&&\inf_{\up}\inf_{\sp}\int_\Omega\,\Biggl[\left(\matrix{-\spp_0 &\epp_0\cr}\right)
\left(\matrix{\ep \cr
              \sp\cr}\right) - \left(\matrix{\pp_0 &\omega\upp_0\cr}\right)
\left(\matrix{\Go\up\cr
              \ppp\cr}\right)\nonumber\\
&&\hbox{\hskip 1in} +\half\left(\matrix{\ep &\sp\cr}\right){\cal C}
\left(\matrix{\ep\cr
              \sp\cr}\right)
+\half\left(\matrix{\omega\up & \ppp\cr}\right){\cal P}
\left(\matrix{\omega\up\cr
               \ppp\cr}\right)\Biggr]\,dx.
\eeqa{18}
Notice that the constitutive relations $\sigma = C e$ and 
$p=i\omega\rho u$ imply the new constitutive relations
\beq \pmatrix{\spp \cr -\epp}=\CC\pmatrix{\ep \cr \sp\cr}, \quad \quad
\pmatrix{\pp \cr \omega\upp}=\CP\pmatrix{\omega\up\cr\ppp\cr}.
\eeq{19}
The fields appearing on the left sides of equations \eq{19}
are linked by the differential constraints
\beq {\rm div}\sigma'' + f'' = \omega p',\quad e'' = \half\{\nabla u'' + (\nabla u'')^T\},
\eeq{19a}
while the fields on the right sides are linked by
\beq {\rm div}\sigma' + f' = -\omega p'', 
\quad e' = \half\{\nabla u' + (\nabla u')^T\}.
\eeq{19b}
Following \citeAPY{Milton:2009:MVP},
we can rewrite \eq{19} as a single equation $G=\CL F$ where
\beq \CL=\pmatrix{\CC & 0 \cr 0 & \CP},\quad
     G=\pmatrix{\spp \cr -\epp \cr \pp \cr \omega\upp}, \quad
     F=\pmatrix{\ep \cr \sp\cr \omega\up\cr\ppp\cr},
\eeq{19c}
and $G$ satisfies the differential constraints \eq{19a} while
$F$ satisfies the differential constraints \eq{19b}, and $\CL$
is positive definite.
If we regard
\eq{19}--\eq{19b} as a primal problem then the variational 
principle for it is
\beq \inf_{F}\int_\Omega\,(-G_0^TF+\half F^T\CL F)\,dx,
\eeq{19d}
which is equivalent to \eq{18}. To check this variational
principle directly, and thus provide an alternative 
derivation of \eq{18}, we compute the first order variation in the integral in 
\eq{19d}:
\beq \int_\GO\,(G-G_0)^T\Gd F\,dx
= \int_{\partial\GO}\left[\Gd\up(\spp-\spp_0)n+(\upp_0-\upp)(\Gd\sp)n\right]\,dS.
\eeq{20}
This vanishes under the stated boundary conditions on the fields. 
It also clearly 
vanishes (and produces a valid variational principle) 
under more general boundary conditions.
For example, although difficult to realize physically, 
one could fix the real part of the displacement
and the real part of the traction at the boundary, so that $\Gd\up=0$
and $(\Gd\sp)n=0$. Then choosing $\upp_0=0$, $\spp_0=0$ and 
$\omega\pp_0=f^{\prime\prime}$ the variational principle \eq{18}
reduces to that of \citeAPY{Milton:2009:MVP}. Alternatively,
by having no constraints on $\Gd\up$ and $(\Gd\sp)n$ one obtains
a variational principle in which the trial fields are not required to satisfy
any boundary condition, and at the minimum $\spp n=\spp_0 n$ and $\upp=\upp_0$ 
on $\partial\GO$.

If one prefers one can rewrite the variational principle in a 
form where only the surface values of $\upp_0$ and $\spp_0 n$ appear.
Using the differential constraints on $\upp_0$, $\epp_0$, $\spp_0$
and $\pp_0$ and integrating by parts, \eq{18} is equivalent to
\beqa
 &&\inf_{\up}\inf_{\sp}\int_\Omega\,\Biggl[-f''u
+\half\left(\matrix{\ep &\sp\cr}\right){\cal C}
\left(\matrix{\ep\cr
              \sp\cr}\right)
+\half\left(\matrix{\omega\up & \ppp\cr}\right){\cal P}
\left(\matrix{\omega\up\cr
               \ppp\cr}\right)\Biggr]\,dx \nonum
&&\hbox{\hskip 1in}+\int_{\partial\GO}\left(-\up\spp_0 n+\upp_0\sp n\right)\,dS.
\eeqa{20.5}
  
When $f=0$ the minimum of the variational principle \eq{18}, or equivalently
\eq{19d} is
\beq
\int_\GO\,\half(G-2G_0)^TF\,dx
= \half\int_{\partial\GO}\left[\up(\spp-2\spp_0)n+(2\upp_0-\upp)\sp n\right]\,dS,
\eeq{21}
and thus depends only on the surface displacements $u$ and $\upp_0$ and surface 
tractions $\sigma n$ and $\spp_0 n$. Thus for any given choice of trial fields
$\underline{u}'$ and $\underline{\Gs}'$ meeting the chosen boundary
conditions, which imply that
\beq (\underline{u}'-\up)(\spp-\spp_0)n=0, \quad (\upp_0-\upp)(\underline{\Gs}'-\sp)n=0,
\quad{\rm on~}\partial\GO,
\eeq{21a}
equations \eq{19} and \eq{21} produce the inequality 
\beqa &&\int_\Omega\,\Biggl[
\half\left(\matrix{\underline{e}' &\underline{\Gs}'\cr}\right){\cal C}
\left(\matrix{\underline{e}'\cr
              \underline{\Gs}'\cr}\right)
+\half\left(\matrix{\omega\underline{u}' & \underline{p}''\cr}\right){\cal P}
\left(\matrix{\omega\underline{u}'\cr
               \underline{p}''\cr}\right)\Biggr]\,dx \nonum
&& \hbox{\hskip 1in} \geq 
\half\int_{\partial\GO}\left[(2\underline{u}'-\up)\spp n 
+\upp(\sp-2\underline{\Gs}')n\right]\,dS,
\eeqa{21b}
in which \eq{21a} has been used to eliminate $\spp_0$ and $\upp_0$. Since
$\spp_0$ and $\upp_0$ do not appear in this equation we may as well 
choose $\spp_0 n=\spp n$ and $\upp_0=\upp$ on $\partial\GO$ in which
case we are free to choose any $\underline{u}'$ and $\underline{\Gs}'$.
The inequality \eq{21b} bounds the 
possible surface fields $u$ and $\sigma n$, i.e. it bounds the 
Dirichlet to Neumann map associated with $\GO$. Alternatively if the
surface fields are known from measurements, then, for each
set of trial fields, the inequality provides
bounds on the moduli $C$ and $\rho$ inside $\GO$ and could potentially be used for
tomography. 

In the limit as $\rpp\to 0$ the last term in the variational principle
\eq{18} will remain finite if and only if 
\beq u'=p''/(\rho'\omega), \eeq{22}
and in this limit the variational principle becomes
\beq
\inf_{\sp}\int_\Omega\,\Biggl[\left(\matrix{-\spp_0 &\epp_0\cr}\right)
\left(\matrix{\ep \cr
              \sp\cr}\right) - \left(\matrix{\pp_0 &\omega\upp_0\cr}\right)
\left(\matrix{\Go\up\cr
              \ppp\cr}\right)
+\half\left(\matrix{\ep &\sp\cr}\right){\cal C}
\left(\matrix{\ep\cr
              \sp\cr}\right)\Biggr]\,dx,
\eeq{23}
where now $p''$ and $u'$ are given by \eq{16} and \eq{22}.

These variational principles are particularly useful if a multiphase
medium occupies $\Omega$ and one of the phases, occupying a subregion
$\Psi$ is lossless, with $C''=0$ and $\rho''=0$. Then assuming
$C'$ and $\rho'$ are positive in $\Psi$ one does not 
have the flexibility to replace $C$ and $\rho$ by $Ce^{i\Gt}$
and $\rho e^{i\Gt}$ since then the imaginary 
parts $Ce^{i\Gt}$ and $\rho e^{i\Gt}$ would lose their
positive semi-definiteness within $\Psi$. In the limit as
$C''\to 0$ in $\Psi$, we see from \eq{11.5} that the last term in \eq{23}
will remain finite if and only if
\beq \sp=\Cp\ep\quad{\rm in~}\Psi.
\eeq{24}
In this limit the variational principle becomes
\beq
\inf_{\sp}\left\{\int_{\Omega}\,\Biggl[\left(\matrix{-\spp_0 &\epp_0\cr}\right)
\left(\matrix{\ep \cr
              \sp\cr}\right) - \left(\matrix{\pp_0 &\omega\upp_0\cr}\right)
\left(\matrix{\Go\up\cr
              \ppp\cr}\right)\Biggr]\,dx
+\int_{\Omega\setminus\Psi}
\half\left(\matrix{\ep &\sp\cr}\right){\cal C}
\left(\matrix{\ep\cr
              \sp\cr}\right)\,dx\right\},
\eeq{25}
in which $p''$ and $u'$ are given by \eq{16} and \eq{22},
and the infimum is over fields $\sp$ such that $u'$
is an exact solution of the elastodynamic equations in $\Psi$,
i.e. \eq{19b}, \eq{22}, and \eq{24}.

\section{Related elastodynamic principle of Hashin--Shtrikman type}
\setcounter{equation}{0}
Recall that the constitutive relations \eq{2} are equivalent, in the case of time-harmonic
fields, to those given in \eq{19}.
Now introduce a ``comparison'' medium, characterized by matrices $\calC_0$, $\CP_0$ (not necessarily obtained from complex tensors corresponding to $C$, $\rho$),
introduce ``polarizations'' $\tpp$, $\etpp$, $\pip$ and $\np$, and define
\begin{equation}
\left(\matrix{\spp \cr
              -\epp\cr}\right) 
  = \calC_0\left(\matrix{\ep\cr
                       \sp\cr}\right)+\left(\matrix{\tpp\cr
                                                    -\etpp\cr}\right),\;\;\;
\left(\matrix{\pp\cr
            \omega\upp\cr}\right)
   = \CP_0\left(\matrix{\omega\up\cr
                         \ppp\cr}\right) +\left(\matrix{\pip\cr
                                                        -\np\cr}\right).
\eeq{1.1}

These relations are consistent with the desired constitutive relations \eq{19}
if the polarizations are chosen so that
\beq
\left(\matrix{\tpp\cr
     -\etpp\cr}\right)= (\CC-\CC_0)\left(\matrix{\ep\cr
                                                 \sp\cr}\right),\;\;\;
\left(\matrix{\pip\cr
      -\np\cr}\right) = (\CP-\CP_0)\left(\matrix{\omega\up\cr
                                          \ppp\cr}\right).
\eeq{1.2a}

With this preparation, note that
\begin{eqnarray}
\half\left(\matrix{\ep &\sp\cr}\right){\cal C}
\left(\matrix{\ep\cr
              \sp\cr}\right) &\approx &\half\left(\matrix{\ep &\sp\cr}\right){\cal C}_0
\left(\matrix{\ep\cr
              \sp\cr}\right)\nonumber\\
&& + \left(\matrix{\ep &\sp\cr}\right)\left(\matrix{\tpp\cr
                                                     -\etpp\cr}\right)
-\half\left(\matrix{\tpp &-\etpp\cr}\right)({\cal C}-{\cal C}_0)^{-1}\left(\matrix{\tpp\cr                                                                                 -\etpp\cr}\right).
\eeqa{1.3}
Similarly,
\begin{eqnarray}
\half\left(\matrix{\omega\up & \ppp\cr}\right)\CP
\left(\matrix{\omega\up\cr
               \ppp\cr}\right) &\approx &
\half\left(\matrix{\omega\up & \ppp\cr}\right)\CP_0\left(\matrix{\omega\up\cr
               \ppp\cr}\right)\nonumber\\
&& \hbox{\hskip -4pt}+ \left(\matrix{\omega\up &           \ppp\cr}\right)\left(\matrix{\pip\cr
                           -\np\cr}\right)
-\half\left(\matrix{\pip &-\np\cr}\right)(\CP-\CP_0)^{-1}
\left(\matrix{\pip\cr                                                                                -\np\cr}\right).
\eeqa{1.4}
Equality is achieved when relations \eq{1.2a} are satisfied.

The Hashin--Shtrikman variant of the minimum principle \eq{17} is now
\begin{eqnarray}
&&\inf_{\up}\inf_{\sp}\int_\Omega\,\Biggl[\left(\matrix{\tpp-\spp_0 &\epp_0-\etpp\cr}\right)
\left(\matrix{\ep \cr
              \sp \cr}\right) \nonum
&&\hbox{\hskip .75in}+\half\left(\matrix{\ep &\sp\cr}\right){\cal C}_0
\left(\matrix{\ep\cr
              \sp\cr}\right)- \half\left(\matrix{\tpp &-\etpp\cr}\right)
              ({\cal C}-{\cal C}_0)^{-1}
\left(\matrix{\tpp\cr
              -\etpp\cr}\right) \nonumber\\
&&\hbox{\hskip .75in}
+\left(\matrix{\pip-\pp_0 & -\np-\omega\upp_0\cr}\right)
\left(\matrix{\omega\up\cr
              \ppp\cr}\right) \nonumber\\
&&\hbox{\hskip .75in}+\half\left(\matrix{\omega\up & \ppp\cr}\right)\CP_0
\left(\matrix{\omega\up\cr
               \ppp\cr}\right)
 -\half\left(\matrix{\pip &-\np\cr}\right)(\CP-\CP_0)^{-1}\left(\matrix{\pip\cr
                                                                                   -\np\cr}\right)\Biggr]\,dx.
\eeqa{1.5}
This produces an approximate solution for any given choice of the ``polarizations'' $\tpp$, $\pip$,
$\etpp$ and $\np$, and the exact solution if these are chosen to make the functional stationary. When ${\cal C}_0$
and $\CP_0$ are chosen so ${\cal C}_0>{\cal C}$ and $\CP_0>\CP$ then the $\approx $ signs in \eq{1.3} and \eq{1.4}
can both be replaced by $\leq$ signs and then one has a minimum principle which involves taking the infimum over
the ``polarizations'' $\tpp$, $\pip$, $\etpp$ and $\np$, as well as an infimum over the fields $\up$ and $\sp$.
If on the other hand  ${\cal C}_0$ and $\CP_0$ are chosen so ${\cal C}_0<{\cal C}$ and $\CP_0<\CP$ then the $\approx $ signs 
can both be replaced by $\geq$ signs and one obtains a saddle principle which involves taking the supremum over
the ``polarizations'' and an infimum over the fields $\up$ and $\sp$. In either case the infimum over the fields $\up$ and $\sp$ 
can be computed in terms of the relevant Green's function and one is left with a minimum principle or a maximum
principle. When the body is infinite in extent the Green's function can be explicitly computed as shown in the 
appendix.

In the notation of \eq{19c} and \eq{19d}, \eq{1.5} becomes
\beq
\inf_F\,\int_\Omega\,\left[(T-G_0)^TF + \half F^T\CL_0 F -\half T^T(\CL-\CL_0)^{-1}T
\right]\,dx,
\eeq{19e}
where
\beq
T=\pmatrix{\tpp \cr -\etpp\cr \pip\cr -\np\cr},
\eeq{19f}
and in this form the principle is easily generalized to acoustics and electromagnetism, as will become
clear in sections 5 and 6.
It is perhaps worth remarking that there is no need for $\CL_0$ to be restricted
to have the block-diagonal structure displayed by $\CL$ in \eq{19c}$_1$; this, however, is not developed further at present.

Developing \eq{19e} a little further, let $F_0$ solve the variational problem for
the comparison medium, {\it i.e.} in the case that $T=0$, and write
\beq
F=F_0+F^\prime.
\eeq{20a}
Any admissible $F^\prime$ must satisfy homogeneous boundary conditions
and hence
\beq
\int_\Omega\,F^{\prime T}(-G_0 + \CL_0F_0)\,dx = 0.
\eeq{20b}
It follows that \eq{19e} can be written
\beq
\inf_{F^\prime}\int_\Omega\,\Bigl(-G_0^TF_0+\half F_0^T\CL_0F_0
+T^T(F_0+F^\prime)+\half F^{\prime T}\CL_0F^\prime
-\half T^T(\CL-\CL_0)^{-1}T\bigr)\,dx.
\eeq{20c}
Formally, if the field $F^\prime$ that attains the infimum is
\beq
F^\prime = -\CH_0 T,
\eeq{20d}
then \eq{20c} becomes
\beq
\int_\Omega\,\Bigl(-G_0^TF_0+\half F_0^T\CL_0 F_0
+ T^T F_0 -\half T^T\left[(\CL-\CL_0)^{-1}+\CH_0\right]T\Bigr)\,dx
\eeq{20e}
and this is stationary with respect to $T$ when
\beq
\left[(\CL-\CL_0)^{-1}+\CH_0\right]T = F_0.
\eeq{20f}
The operator $\CH_0$ relevant to an infinite
medium is given explicitly in the appendix.

\section{Variational principles for acoustics}
\setcounter{equation}{0}
For acoustics at fixed frequency in the presence of a body force $f$ the 
fields satisfy the differential constraints
\beq -\Grad P +f=i \Go p, \quad h=\Div v,
\eeq{2.1}
where $P$ is the pressure, $p$ the momentum, $v$ the velocity, and  
$h$ is defined to be the divergence of the velocity. They also
satisfy the constitutive relations
\beq  h=-i \Go k P, \quad v=r p, \eeq{2.2}
where $k=1/\Gk$ is the compressibility, the inverse of the bulk modulus
$\Gk$, which is  complex with $k''<0$, and $r=\rho^{-1}$ is the inverse density (which we allow to be matrix
valued and complex with $r''>0$). Introducing the positive
definite matrices
\beq \CK  =  -\pmatrix{k''+(k')^2/k'' & 
                   -k'/k'' \cr 
                   -k'/k'' & 1/k''}, \quad
\CR  =  \pmatrix{r''+r'(r'')^{-1}r' & -r'(r'')^{-1} \cr 
                   -(r'')^{-1}r' & (r'')^{-1}},
\eeq{2.3}
the constitutive relations take the new form
\beq \pmatrix{h' \cr -\omega P''}=\CK\pmatrix{-\omega P'\cr h''\cr}, \quad \quad
 \pmatrix{-v' \cr p'}=\CR\pmatrix{p'' \cr v''\cr},
\eeq{2.4}
which can be expressed as the single equation $G=\CL F$ with
\beq \CL=\pmatrix{\CK & 0 \cr 0 & \CR}, \quad
G=\pmatrix{h' \cr -\omega P'' \cr -\Go v' \cr \Go p'}, \quad
F=\pmatrix{-\omega P'\cr h''\cr \Go p'' \cr \Go v''\cr},
\eeq{2.4aa}
where the fields appearing in $G$ 
are linked by the differential constraints
\beq \quad h'=\Div v', \quad -\Grad P'' +f''=\Go p', \eeq{2.4a}
while the fields appearing in $F$ are linked by
\beq -\Grad P' +f'=-\Go p'', \quad h''=\Div v''.
\eeq{2.5}
(The second equation in \eq{2.4} was written with the real components of the
fields on the left, rather than on the right, to ensure that the differential
constraints couple fields on the same sides of \eq{2.4}.) 
Based on \eq{19d}, 
we assert that the form of the variational principle for acoustics is
\beqa
&&\inf_{v''}\inf_{P'}\int_\Omega\,\Biggl[\left(\matrix{-h'_0 &\omega P''_0\cr}\right)
\left(\matrix{-\Go P'\cr h''\cr}\right)
+\Go^2\left(\matrix{v'_0 &-p'_0}\right)
\left(\matrix{p'' \cr
              v''\cr}\right) \nonum
&&\hbox{\hskip 1in} 
+\half\left(\matrix{-\omega P' & h''\cr}\right){\cal K}
\left(\matrix{-\omega P'\cr
               h''\cr}\right)
+\half\Go^2\left(\matrix{p'' & v''\cr}\right){\cal R}
\left(\matrix{p''\cr
              v''\cr}\right)\Biggr]\,dx,
\eeqa{2.6}
in which $h''$ and $p''$ are given by \eq{2.5}, and $h'_0$, $P''_0$, $v'_0$, and $p'_0$
satisfy the differential constraints \eq{2.4a}. To check this we see that
the first order variation
\beqa &&\int_\Omega\,\Biggl[\left(\matrix{h'-h'_0 &\omega P''_0-\omega P''\cr}\right)
\left(\matrix{-\Go \Gd P'\cr \Gd h''\cr}\right)+
\Go^2\left(\matrix{v'_0-v' &p'-p'_0\cr}\right)
\left(\matrix{\Gd p'' \cr
              \Gd v''\cr}\right)\Biggr]\,dx \nonum
&& \hbox{\hskip 1in} 
=\Go\int_{\partial\GO}\left[\Gd P'(v'_0-v')n+(P''_0-P'')(\Gd v'')n\right]\,dS,
\eeqa{2.7}
only depends on boundary terms. (Note the need for the factor $\Go^2$ in
\eq{2.7} which explains the need for the additional factors of 
$\Go$ in the last two components of $G$ and $F$ in \eq{2.4aa}).

The condition that the boundary integral vanishes for 
all $\Gd P'$ and $\Gd v''$ that 
are permitted according to which components of $P'$ and $v''n$ are prescribed
on $\partial\GO$, forces 
the complementary components of $(v'_0-v')n$ and $P''_0-P''$ to be zero, thus
fixing these components of $v'n$ and $P''$. In the particular case
when $f=0$ and all components of $P'$ and $v''n$ are prescribed at the boundary,
then by choosing $P''_0=0$, $p'_0=0$, and $v'_0=0$, these variational principles reduce to those of \citeAPY{Milton:2009:MVP}.

If one prefers one can rewrite \eq{2.6} in a form where only the surface 
values of $P''_0$ and $v'_0 n$ appear. Using integration by parts the 
variational principle is equivalent to
\beqa
&&\inf_{v''}\inf_{P'}\int_\Omega\,\Biggl[-\Go f''v''
+\half\left(\matrix{-\omega P' & h''\cr}\right){\cal K}
\left(\matrix{-\omega P'\cr
               h''\cr}\right)
+\half\Go^2\left(\matrix{p'' & v''\cr}\right){\cal R}
\left(\matrix{p''\cr
              v''\cr}\right)\Biggr]\,dx \nonum
&&\hbox{\hskip 1in} 
+\Go\int_{\partial\GO}\left(P'v'_0 n+P''_0 v''n\right)\,dS.
\eeqa{2.7a}

When $f=0$ the minimum value of \eq{2.6} or \eq{2.7a} 
only depends on the boundary fields, and is
\beq \half \Go\int_{\partial\GO}\left[P'(2v'_0-v')n+(2P''_0-P'')v''n\right]\,dS.
\eeq{2.8}
Thus knowledge of this boundary data will, for any given choice of trial fields
$\underline{v}''$ and $\underline{P}'$,
give constraints 
\beqa &&\int_\Omega\,\Biggl[
\half\left(\matrix{-\omega \underline{P}' & \underline{h}''\cr}\right){\cal K}
\left(\matrix{-\omega \underline{P}'\cr
               \underline{h}''\cr}\right)
+\half\Go^2\left(\matrix{\underline{p}'' & \underline{v}''\cr}\right){\cal R}
\left(\matrix{\underline{p}''\cr
              \underline{v}''\cr}\right)\Biggr]\,dx \nonum
&&\hbox{\hskip 1in} \geq
\half \Go\int_{\partial\GO}\left[(P'-2\underline{P}')v'n
+P''(v''-2\underline{v}'')n\right]\,dS
\eeqa{2.8a}
on the possible values of the moduli $\kappa$ and $\rho$ inside $\GO$,
and thus may be useful for tomography. 

In the limit as $r''\to 0$, i.e. as $\rho''\to 0$ the last term in \eq{2.6} will
be minimized when 
\beq v''=r'p'', \eeq{2.9}
and is infinite otherwise. In this limit the variational principle reduces 
to 
\beq
\inf_{P'}\int_\Omega\,\Biggl[\left(\matrix{-h'_0 &\omega P''_0\cr}\right)
\left(\matrix{-\Go P'\cr h''\cr}\right)
+\Go^2\left(\matrix{v'_0 &-p'_0}\right)
\left(\matrix{p'' \cr
              v''\cr}\right) 
+\half\left(\matrix{-\omega P' & h''\cr}\right){\cal K}
\left(\matrix{-\omega P'\cr
               h''\cr}\right)\Biggr]\,dx,
\eeq{2.10}
where $h''$ and $p''$ are given by \eq{2.5}, while $v''$ is given by \eq{2.9}.

\section{Variational principles for electrodynamics}
\setcounter{equation}{0}
For electromagnetism it is more conventional to have fields which are multiplied
by $e^{-i\Go t}$, rather than $e^{i\Go t}$, so that the permittivity tensor $\Gve$ and
permeability tensor $\Gm$ have positive definite imaginary parts. We now adopt this convention.
Then the fields in Maxwell's equations at fixed frequency $\Go$ satisfy the
differential constraints
\beq \Curl E=i\Go B,\quad \Curl H=-i\Go D+j, \eeq{3.8}
and the constitutive equations
\beq D=\Gve E,\quad H= m B, \eeq{3.9}
where $m=\Gm^{-1}$.
Introducing the positive definite matrices
\beq \CE  =  \pmatrix{\Gve''+\Gve'(\Gve'')^{-1}\Gve' & -\Gve'(\Gve'')^{-1} \cr 
                   -(\Gve'')^{-1}\Gve' & (\Gve'')^{-1}}, \quad
\CM  =  -\pmatrix{m''+m'(m'')^{-1}m' & 
                   -m'(m'')^{-1} \cr 
                   -(m'')^{-1}m' & (m'')^{-1}},
\eeq{3.10}
the constitutive relations take the new form 
\beq \pmatrix{D'' \cr - E''}=\CE\pmatrix{E' \cr D'}, \quad
\pmatrix{H' \cr -B'}=\CM\pmatrix{B''\cr H''}
\eeq{3.11}
which can be expressed as the single equation $G=\CL F$ with
\beq \CL=\pmatrix{\CE & 0 \cr 0 & \CM}, \quad
G=\pmatrix{D'' \cr -E'' \cr H' \cr -B'}, \quad
F=\pmatrix{E'\cr D'\cr B'' \cr H''\cr},
\eeq{3.11a}
where the fields appearing in $G$ 
are linked by the differential constraints
\beq \Curl H'-j'=\Go D'', \quad \Curl E''=\Go B', \eeq{3.12}
while the fields appearing in $F$ are linked by 
\beq   \Curl E'=-\Go B'', \quad \Curl H''-j''=-\Go D'. \eeq{3.13}
Based on \eq{19d} 
we state that the form of the variational principle for electromagnetism is
\beqa
&&\inf_{E'}\inf_{H''}\int_\Omega\,\Biggl[\left(\matrix{-D''_0 & E''_0\cr}\right)
\pmatrix{E' \cr D'}
+\left(\matrix{-H'_0 & B'_0}\right)
\pmatrix{B''\cr H''} \nonum
&&\hbox{\hskip 1in} 
+\half\pmatrix{E' & D'}{\cal E}\pmatrix{E' \cr D'}
+\half\pmatrix{B'' & H''}{\cal M}\pmatrix{B''\cr H''}\Biggr]\,dx,
\eeqa{3.14}
where $D'$ and $B''$ are given by \eq{3.13}, while $D''_0$, $E''_0$,
$H'_0$ and $B'_0$ satisfy the differential constraints \eq{3.12}. 
To check this we use the identities
\beqa &&\Div(E' \times H')
=H'\cdot(\Curl E')-E'\cdot(\Curl H')
=-\Go(H'B''+E'D'')-j'E', \nonum
&& \Div(E'' \times H'')
=H''\cdot(\Curl E'')-E''\cdot(\Curl H'')
=\Go(H''B'+E''D')-j''E'',
\eeqa{3.15}
which hold for any fields satisfying the differential constraints \eq{3.12}
and \eq{3.13}. Then we
see that the first order variation can be expressed in terms of boundary fields: 
\beqa &&\int_\Omega\,\Biggl[\left(\matrix{D''-D''_0 & E''_0-E''\cr}\right)
\pmatrix{\Gd E' \cr \Gd D'}
+\left(\matrix{H'-H'_0 & B'_0-B'}\right)
\pmatrix{\Gd B''\cr \Gd H''}\Biggr]\,dx,
\nonum
&& \hbox{\hskip 1in} =\frac{1}{\Go}\int_\Omega\,\Biggl[
+\Div((H'-H'_0)\times \Gd E')+\Div((E''_0-E'')\times\Gd H'')\Biggr]\,dx,
\nonum
&& \hbox{\hskip 1in} =\frac{1}{\Go} \int_{\partial\GO}\left[
((H'-H'_0)\times \Gd E')n+((E''_0-E'')\times\Gd H'')n\right]\,dS.
\eeqa{3.16}
Note that this surface integral only depends on the tangential
components of the fields $H'-H'_0$, $\Gd E'$, $E''_0-E''$, and $\Gd H''$.
The condition that the boundary integral vanishes for all $\Gd E'$ and $\Gd H''$ that 
are permitted according to which tangential components of $E'$ and $H''$ 
are prescribed, forces 
the complementary tangential components of $H'-H'_0$ and $E''_0-E''$ to be zero, thus
fixing these tangential components of $H'$ and $E''$. Again
these variational principles reduce to those of \citeAPY{Milton:2009:MVP}
when the tangential components of $E'$ and $H''$ are fully prescribed.

If one prefers one can rewrite \eq{3.14} in a form where only the surface values
of $H'_0$ and $E''_0$ appear. Using integration by parts the variational principle
is equivalent to
\beqa && \inf_{E'}\inf_{H''}\int_\Omega\,\Biggl[j'E'/\Go
+\half\pmatrix{E' & D'}{\cal E}\pmatrix{E' \cr D'}
+\half\pmatrix{B'' & H''}{\cal M}\pmatrix{B''\cr H''}\Biggr]\,dx \nonum
&&\hbox{\hskip 1in}+\int_{\partial\GO}\left[
-(H'_0\times E')n+(E''_0\times H'')n\right]\,dS.
\eeqa{3.16a}

When $j=0$ the minimum value of \eq{3.14} or \eq{3.16a} is
\beq \frac{1}{2\Go}\int_{\partial\GO}\left[
((H'-2H'_0)\times E')n+((2E''_0-E'')\times H'')n\right]\,dS.
\eeq{3.17}
Consequently for any given choice of trial fields $\underline{E}'$
and $\underline{H}''$ we have the inequality
\beqa  && \int_\Omega\,\Biggl[
\half\pmatrix{\underline{E}' & \underline{D}'}{\cal E}
\pmatrix{\underline{E}' \cr \underline{D}'}
+\half\pmatrix{\underline{B}'' & \underline{H}''}{\cal M}
\pmatrix{\underline{B}''\cr \underline{H}''}\Biggr]\,dx \nonum
&&\hbox{\hskip 1in} \geq
\frac{1}{2\Go}\int_{\partial\GO}\left[(H'\times(2\underline{E}'-E'))n+
(E''\times(H''-2\underline{H}''))n \right]\,dS,
\eeqa{3.17a}
which could be useful for tomography if we have measurements of the
boundary values of the fields.

In the limit as $\mu''\to 0$, i.e. as $m''\to 0$, the last term in the
\eq{3.14} will be minimized when
\beq H''=m' B'', \eeq{3.18}
and is infinite otherwise. In this limit the variational principle reduces to
\beq
\inf_{E'}\int_\Omega\,\Biggl[\left(\matrix{-D''_0 & E''_0\cr}\right)
\pmatrix{E' \cr D'}
+\left(\matrix{-H'_0 & B'_0}\right)
\pmatrix{B''\cr H''}  
+\half\pmatrix{E' & D'}{\cal E}\pmatrix{E' \cr D'}
\Biggr]\,dx,
\eeq{3.19}
where $D'$ and $B''$ are given by \eq{3.13}, while $H''$ is given by \eq{3.18}. 

\section*{Appendix: Green's function for infinite comparison medium}
\renewcommand\theequation{A.\arabic{equation}}
\setcounter{equation}{0}

The Hashin--Shtrikman variational functional is rendered stationary (with respect
to the fields $\ep,\, \sp,\, \omega\up,\, \ppp$) when $\epp$ is the strain field
associated with the displacement $\upp$ and the equation of motion
\beq
{\rm div}\left(\matrix{\spp\cr
                       \sp\cr}\right)+\left(\matrix{\fpp\cr
                                                     \fp\cr}\right)
          =\omega\left(\matrix{\pp\cr
                               -\ppp\cr}\right)
\label{A.1}\end{equation}
is satisfied, together with the relations \eq{1.1}.

Motivated by the form of relations \eq{9.5}, define
\begin{equation}
\calD_0 = \left(\matrix{D_2 & D_1\cr
                        D_1^T & -D_3\cr}\right)\;\hbox{ and }\;
\calQ_0 = \left(\matrix{Q_2 & Q_1\cr
                        Q_1^T & -Q_3\cr}\right)
\label{A.0}\end{equation}
in which $D_2$, $D_3$, $Q_2$ and $Q_3$ are symmetric real matrices
and, in terms of these and the real matrices $D_1$ and $Q_1$,
\begin{equation}
\calC_0 = \left(\matrix{D_2 + D_1 D_3^{-1}D_1^T & -D_1D_3^{-1}\cr
                       -D_3^{-1}D_1^T         & D_3^{-1}\cr}\right)
\;\hbox{ and }\;
\calP_0 = -\left(\matrix{Q_2 + Q_1 Q_3^{-1}Q_1^T & -Q_1Q_3^{-1}\cr
                       -Q_3^{-1}Q_1^T         & Q_3^{-1}\cr}\right).
\label{A.0c}\end{equation}
The relations \eq{1.1} may then be given in the equivalent form
\begin{equation}
\left(\matrix{\spp-\tpp\cr
              \sp\cr}\right) = \calD_0\left(\matrix{\ep\cr
                                                    \epp-\etpp\cr}\right),\;\;\;
\left(\matrix{\pp-\pip\cr
              -\ppp\cr}\right) = -\calQ_0\left(\matrix{\omega\up\cr
                                                       \omega\upp+\np\cr}\right).
\label{A.0a}\end{equation}

The quadratic forms associated with $\calC_0$ and $\calP_0$ are positive-definite if
$D_2,D_3,-Q_2$ and $-Q_3$ are positive-definite, as will henceforth be assumed.

The equation of motion \eq{A.1} now gives 
\beq
{\rm div}(\CD_0 \nabla U) + \omega^2\CQ_0 U + \tilde f = 0,
\label{A.6}
\end{equation}
where
\beq
U = \left(\matrix{\up\cr
                  \upp\cr}\right)\;\hbox{ and }\;
\tilde f=\left(\matrix{\fpp+{\rm div}(\tpp-D_1\etpp)+\omega (Q_1\np-\pip)\cr
                \fp+{\rm div}(D_3\etpp)-\omega Q_3\np\cr}\right).
\label{A.7}\end{equation}

The required Green's function $\CG_0$ is that associated with equation \eq{A.6}. It
can be constructed by a method introduced for the ``ordinary'' elastodynamic Green's function by \citeAPY{Willis:1980:PA1}. First, select $\tilde f(x)
=\tilde f_0\delta(x)$, with $\tilde f_0$ constant,
and note the plane-wave decomposition of the three-dimensional delta-function
\beq
\delta(x)  = -\frac{1}{8\pi^2}\int_{\vert\xi\vert =1}\,
\delta^{\prime\prime}(\xi\cdot x)\,dS.
\label{A.8}\end{equation}
This motivates first solving (A.6), with $\tilde f$ replaced by
$\tilde f_0\delta(\xi\cdot x)$, where $\vert\xi\vert$=1. The
solution, $U^\xi$ say, depends on $x$ only in the combination $s=\xi\cdot x$; thus,
it satisfies the system of ODE's
\beq
\CL_0\frac{d^2 U^\xi}{ds^2}+\omega^2\CQ_0 U^\xi + \tilde f_0\delta (s) = 0,
\label{A.9}\end{equation}
where
\beq
\CL_0 = \left(\matrix{\xi^T D_2\xi &\xi^T D_1\xi\cr
                      \xi^T D_1^T\xi & -\xi^T D_3\xi\cr}\right).
\label{A.10}\end{equation}
The system can be diagonalized by reference to the eigenvalue problem
\beq
\CL_0 U_N = c_N^2 \CQ_0 U_N\;\;\;(N=1,2,\cdots 6).
\label{A.11}\end{equation}
Although the matrices are real and symmetric, they are not positive-definite and the
eigenvalues may be real or they may occur in complex conjugate pairs. Suppose first
that \eq{A.11} is satisfied for some real $c_N^2$ and (hence real) $U_N$, having the form shown in equation (13)$_1$: say $U_N = (\up_N,\upp_N)^T$, where $\up_N$ and $\upp_N$ are 3-vectors. 
The 3-vector $\upp_N$ can be eliminated from \eq{A.11}, to give
\begin{equation}
\{[(\xi D_2\xi)-c_N^2 Q_2]+ [(\xi D_1\xi)-c_N^2 Q_1][(\xi D_3\xi)-c_N^2
 Q_3]^{-1}[(\xi D_1\xi)-c_N^2 Q_1]^T\}\up_N = 0
\end{equation}
and hence
\begin{equation}
(\up_N)^T\{[(\xi D_2\xi)-c_N^2 Q_2]+ [(\xi D_1\xi)
-c_N^2 Q_1][(\xi D_3\xi)-c_N^2 Q_3]^{-1}[(\xi D_1\xi)-c_N^2 Q_1]^T\}\up_N = 0.
\label{16a}\end{equation}
Recalling the definiteness of $D_2,D_3,-Q_2$ and $-Q_3$, it follows that the expression on the left
side of \eq{16a} cannot be zero if $c_N^2$ is positive. Hence, any real eigenvalue must be negative.
Now write
\begin{equation}
c_N^2 =(\cp_N + i\cpp_N)^2,\;\;N=1,\cdots 6.
\end{equation}
Since $\cpp_N$ cannot be zero, it is possible to ensure that $\cpp_N >0$ for all $N$. If,
furthermore, $c_N^2$ is complex, then $c_M^2 = \overline{c_N^2}$ and $U_M =
\overline{U_N}$ for some $M$, and correspondingly $c_M = -\overline{c_N}$.
The eigenvectors are normalized so that
\begin{equation}
U_N^T\calQ_0 U_M = \delta_{NM}.
\end{equation}
Setting
\beq U^\xi(s)=\sum_{M=1}^6\phi_M(s)U_M,
\label{A.11a}\end{equation}
it follows that
\beq
(\cp_N+i\cpp_N)^2\frac{d^2 \phi_N}{ds^2}+\omega^2 \phi_N 
+ U_N^T \tilde f_0\delta(s) = 0.
\label{A.15}\end{equation}
The solution that decays to zero as $\vert s\vert \to \infty$ is
\beq
\phi_N(s) = \frac{U_N^T \tilde f_0\,\exp{\left(\frac{-i\omega\vert s\vert}
{(\cp_N+i\cpp_N)}\right)}}
{2i\omega(\cp_N+i\cpp_N)}.
\label{A.16}\end{equation}
Having now determined $U^\xi(s)$, the solution
$U(x)$ of (\ref{A.6}), with $\tilde f(x) = \tilde f_0\delta(x)$, follows by making the superposition implied by (\ref{A.8}):
\beq
U(x) = \CG_0(x)\tilde f_0,
\label{A.17}\end{equation}
where
\beq
\CG_0(x) = \frac{1}{8\pi^2}\left\{\sum_{N=1}^6\int_{\vert\xi\vert=1}\,
\frac{U_NU_N^T}{(\cp_N+i\cpp_N)^2}\left[\delta(\xi\cdot x)
-\frac{i\omega}{2(\cp_N+i\cpp_N)}\exp\left(
\frac{-i\omega\vert\xi\cdot x\vert}{(\cp_N+i\cpp_N)}\right)\right]\,dS\right\}.
\label{A.18}\end{equation}
If, for any $N$, the eigenvalue and eigenvector are real, then $\cp_N=0$ and the contribution
to the sum is real. If the eigenvalue is complex, it follows from the properties established above
that some other $M$ contributes the complex conjugate expression. Thus, $\CG_0$ is real.

Now for some general $\tilde f(x)$, and in particular the $\tilde f(x)$ given by \eq{A.7}, the solution is
\beq U(x)=\int \CG_0(x-y)\tilde f(y)\,dy
\label{A.19}\end{equation}
which gives us $\up$ and $\upp$ and hence $\ep$ and $\epp$, and from \eq{A.0a} we get
\beq \sp=D_1^T\ep+D_3(\etpp-\epp),\quad\quad \ppp=Q_1^T\omega \up-Q_3(\omega\upp+\np).
\label{A.20}\end{equation}
The functions $\up$, $\sp$, $\ep$, and $\ppp$ thus obtained in terms
of the polarization fields are then substituted back in \eq{1.5}.
The resulting functional is a minimization (or maximization) principle
for the polarization fields when ${\cal C}_0$ and $\CP_0$ are chosen
so both ${\cal C}_0-{\cal C}$ and $\CP_0-\CP$ are positive definite
(respectively, negative definite).

This appendix is concluded by
recording the explicit form of the operator $\CH_0$ introduced in \eq{20d}.
For this purpose, it is helpful to define
\beq
\CE_{ijk} = \half(\delta_{ik}\partial_j+\delta_{jk}\partial_i)
\eeq{A.21}
so that the strain $e$ associated with displacement $u$ is $e=\CE u$, and also
to define
\beq
\CE^\dag_{kji} = \CE_{ijk},
\eeq{A.22}
but with the partial derivatives acting {\it backwards}, with respect to the
immediately preceding variable, so that, for example,
\beq
(G\CE^\dag)_{ikl} = \frac{1}{2}\left\{\frac{\partial G_{ik}(x,y)}{\partial y_l}
+\frac{\partial G_{il}(x,y)}{\partial y_k}\right\}.
\eeq{A.23}
We also write
\beq
\CG_0 = \left(\matrix{G_2 & G_1\cr
                    G_1^T & -G_3\cr}\right).
\eeq{A.24}
In terms of the representation \eq{20d} employing the polarization $T$ as defined in
\eq{19f}, it follows from \eq{A.6}, \eq{A.19} and \eq{A.20} that
\beq
{\scriptsize
\hbox{\hskip -.25in}
\CH_0=\left[\matrix{\CE G_2\CE^\dag & \CE(G_2\CE^\dag D_1-G_1\CE^\dag D_3) &
     \omega\CE G_2 & \omega\CE (G_2 Q_1-G_1 Q_3)\cr
     &&&\cr
     (D_1^T\CE G_2-D_3\CE G_1^T)\CE^\dag & \bigl\{D_3 + D_1^T\CE(G_2\CE^\dag
      D_1-G_1\CE^\dag D_3) &\omega(D_1^T\CE G_2 - D_3\CE G_1^T) &
      \omega\bigl\{D_1^T\CE(G_2Q_1-G_1Q_3)\cr
       & - D_3\CE(G_1^T\CE^\dag D_1+G_3\CE^\dag D_3)\bigr\} & &
       -D_3\CE(G_1^T Q_1+G_3 Q_3)\bigr\}\cr
       &&&\cr
 \omega G_2\CE^\dag & \omega(G_2\CE^\dag D_1 - G_1\CE^\dag D_3) & \omega^2 G_2 & \omega^2(G_2 Q_1-G_1 Q_3)\cr
 &&&\cr
 \omega(Q_1^T G_2 -Q_3 G_1^T)\CE^\dag &
 \omega\bigl\{Q_1^T(G_2\CE^\dag D_1 - G_1\CE^\dag D_3) &
  \omega^2(Q_1^TG_2-Q_3G_1^T)& -\bigl\{Q_3 -\omega^2Q_1^T(G_2Q_1-G_1Q_3)\cr
            &-Q_3(G_1^T\CE^\dag D_1+G_3\CE^\dag D_3)\bigr\} &&+\omega^2Q_3(G_1^TQ_1+G_3Q_3)
            \bigr\}\cr
 }\right].
 }
\eeq{A.25}
The symmetry of $\CH_0$ is consistent with its association with the variational
formulation for the comparison medium.

\section*{Acknowledgements}
The authors are grateful for support from the
National Science Foundation through grant DMS-070978.

\bibliography{/u/ma/milton/tcbook,/u/ma/milton/newref}

\end{document}